\documentclass{article}
\usepackage{graphicx}
\usepackage{amsmath}
\usepackage{amsfonts}
\usepackage{amssymb}

\newcommand{\be}{\begin{equation}}
\newcommand{\ee}{\end{equation}}
\newcommand{\tr}{{\rm Tr}}
\newcommand{\Wg}{{\rm Wg}}
\newcommand{\Z}{\mathcal{Z}}
\newcommand{\U}{\mathcal{U}}
\newcommand{\iv}{\overset{{}_{\leftrightarrow}}{i}}
\newcommand{\jv}{\overset{{}_{\leftrightarrow}}{j}}
\newcommand{\av}{\overset{{}_{\leftrightarrow}}{a}}
\newcommand{\bv}{\overset{{}_{\leftrightarrow}}{b}}
\newcommand{\cv}{\overset{{}_{\leftrightarrow}}{c}}

\begin{document}

\title{Energy-dependent correlations in the $S$-matrix of chaotic systems}
\author{Marcel Novaes\\ Instituto de F\'isica, Universidade Federal de Uberl\^andia\\ Av. Jo\~ao
Naves de \'Avila 2121, Uberl\^andia, MG, 38408-100, Brazil}
\maketitle
\begin{abstract}
The $M$-dimensional unitary matrix $S(E)$, which describes scattering of waves, is a strongly fluctuating function of the energy for complex systems such as ballistic cavities, whose geometry induces chaotic ray dynamics. Its statistical behaviour can be expressed by means of correlation functions of the kind $\left \langle S_{ij}(E+\epsilon)S^\dag_{pq}(E-\epsilon)\right\rangle$, which have been much studied within the random matrix approach. In this work, we consider correlations involving an arbitrary number of matrix elements and express them as infinite series in $1/M$, whose coefficients are rational functions of $\epsilon$. From a mathematical point of view, this may be seen as a generalization of the Weingarten functions of circular ensembles.
\end{abstract}

\section{Introduction}

Scattering of waves of given frequency can be described by the so-called $S-$matrix, which connects incoming waves to outgoing waves and is usually treated as a function of energy, $S(E)$. We consider a scattering region connected to the outside world by $M$ perfectly transparent channels, so that $S$ is a $M$-dimensional matrix, unitary as a consequence of energy conservation. When the system has time-reversal symmetry (TRS), $S$ must also be symmetric.

We are interested in systems for which the ray dynamics is strongly chaotic. This situation can be realized in microwave scattering in metallic cavities\cite{micro1,micro2,micro3,micro4,micro5,micro6} and in electron scattering in condensed matter systems,\cite{elec1,elec2,elec3,elec4,elec5} among other possibilities. We assume there is a well defined decay rate for the ray dynamics, $\Gamma$; this means that the total energy inside the system decays exponentially in time as $e^{-\Gamma t}$. In classical dynamics language, the average amount of time spent inside the cavity by a particle injected at random is $\tau_D = 1/\Gamma$, the ``dwell time''.

When the wavelength in question is much smaller than the typical linear dimension of the scattering region, the matrix elements of $S$ are strongly oscillating functions of the energy. In this situation, trying to compute them accurately is too costly and a statistical approach is in order. One possibility is to replace $S$ by a random matrix:\cite{haake,beenakker,mehta} for broken TRS it is taken to be uniformly distributed in the Circular Unitary Ensemble (CUE), which is just the unitary group with Haar measure, while for intact TRS it is taken to be uniformly distributed in the Circular Orthogonal Ensemble (COE), which is just the space of unitary symmetric matrices with appropriate measure. An `ergoditicy hypothesis' then replaces energy averages with ensemble averages.

This random matrix theory (RMT) approach has had impressive success and is very popular in the calculation of transport statistics.\cite{prb78,prb80,prb80mac,prl101} In particular, fixed-energy correlation functions of matrix elements are well known for the CUE,\cite{samuel,BB,esposti,collins} and for the COE.\cite{coe,matsucompact} Such group-theoretical averages are computed in terms of so-called Weingarten functions, for which there are explicit expressions involving characters and zonal spherical functions associated with symmetric groups. We review these results in Section II.

A new level of complexity arises if one wishes to consider energy-correlations between the elements of $S$. Within RMT, this is usually treated using the so-called Heidelberg formulation which, instead of addressing the statistics of $S$ directly, introduces randomness in the system's Hamiltonian.\cite{micro2,sigma1,sigma2,sigma3} This has been used to compute correlations involving up to four matrix elements,\cite{new1,new2,new3,new4} and leads to exact results which are not quite explicit, since some difficult integrals remain to be evaluated. Asymptotic expansions in the parameter $1/M$ are available for the simplest case of only two elements.\cite{new5}

In this work we consider correlations of the form \be\label{corr1} C_1(\vec{i},\vec{j},\vec{p},\vec{q},M,\epsilon)=\left\langle \prod_{k=1}^n S_{i_kj_k}(E_+)S^\dag_{p_kq_k}(E_-)\right\rangle_{\rm CUE(M)}\ee for broken TRS (here $\vec{i}=(i_1,\ldots,i_n)$), and 
\be\label{corr2} C_2(\iv,\jv,M,\epsilon)=\left\langle \prod_{k=1}^n S_{i_ki_{-k}}(E_+)S^\dag_{j_kj_{-k}}(E_-)\right\rangle_{\rm COE(M)}\ee for intact TRS (here $\iv=(i_{-n},\ldots,i_{-1},i_1,\ldots,i_n)$), where \be E_\pm=E\pm\frac{\epsilon \hbar}{2\tau_D}\ee and the averages are taken over $E$. We are able to express these functions as power series in $1/M$, whose coefficients depend rationally of $\epsilon$. This can be seen as a one-parameter generalization of CUE and COE Weingarten functions, and our formulas also involve characters and zonal spherical functions associated with symmetric groups.

Instead of treating $S(E)$ as a random matrix, we employ a semiclassical approximation in which the elements of $S$ are written as infinite sums over scattering rays.\cite{jalabert,sieber,muller} It was used by Kuipers and Sieber\cite{KS1,KS2} to compute two-element correlations and then by Berkolaiko and Kuipers\cite{BK1} for calculations involving an arbitrary number of elements. These results, restricted to leading order $1/M$ expansions, were later used to obtain the density of states of chaotic Andreev billiards.\cite{andreev1,andreev2} The first few $1/M$ corrections have also been treated.\cite{BK2} In a recent development,\cite{previous} restricted to broken TRS, some correlation functions were expressed as power series in $\epsilon$ with coefficients that are rational functions of $M$. 

In the present work, we follow recent advances in semiclassical theory\cite{novaes2,novaes1} in order to formulate correlation functions in terms of auxiliary matrix integrals. These integrals are then computed using the machinery of Jack polynomials and Jack characters. 

In Section II, we discuss correlation functions for fixed-energy, also known as Weingarten functions of CUE and COE. In Section III, we review the semiclassical approximation. In Sections IV and V we present our results for broken TRS and intact TRS, respectively. In Section VI we discuss a particular kind of correlation functions which involves only traces. In an appendix we review needed results from the theory of permutations groups, Jack polynomials and Jack characters (this material is covered, for example, in the books by MacDonald\cite{macdonald} and Sagan\cite{sagan}).

\section{Fixed-energy correlations: Weingarten functions}

For systems with broken time-reversal symmetry (TRS), the matrix $S(E)$ may be modelled as a random matrix uniformly distributed in the Circular Unitary Ensemble (CUE), which is nothing but the unitary group $\mathcal{U}(M)$ with normalized Haar measure. This is true for any fixed value of the energy $E$.

In this case it is known that the correlation function $\left\langle \prod_{k=1}^n S_{i_kj_k}S^\dag_{p_kq_k}\right\rangle$ will be different from zero if and only if the $q$ labels are a permutation of the $i$ labels, and the $p$ labels are a permutation of the $j$ labels. Namely, 
\be\label{e0U} \left\langle \prod_{k=1}^n S_{i_kj_k}S^\dag_{p_kq_k}\right\rangle_{\rm CUE(M)} = \sum_{\sigma,\tau \in S_n} \Wg^{(1)}_M(\sigma^{-1}\tau)\delta_\tau[\vec{q},\vec{i}]\delta_\sigma[\vec{p},\vec{j}].\ee The coefficient ${\rm Wg}^{(1)}_M$, a function on the permutation group $S_n$, is called Weingarten function of the CUE. For example, when the argument is the transposition $(12)$ we have \be\label{ex1} \Wg^{(1)}_M((12))=\langle
S_{1,1}S^\dag_{1,2}S_{2,2}S^\dag_{2,1}\rangle_{\rm CUE(M)}=\frac{-1}{(M-1)M(M+1)},\ee 

The general case has a character expansion given by \cite{samuel,esposti,collins}
\be\label{oldweing} \Wg^{(1)}_M(\pi)=\frac{1}{n!}\sum_{\lambda\vdash n}\frac{d_\lambda}{[M]^\lambda_{(1)}}\chi_\lambda(\pi).\ee
Here $d_\lambda$ is the dimension of the irreducible representation of the permutation group labelled by $\lambda$, $\chi$ are the irreducible characters of that group, and $[M]^\lambda_{(1)}$ is a generalization of the raising factorial. These concepts are reviewed in the appendix.

For systems with intact TRS, the unitary symmetric matrix $S(E)$ may be modelled as a random matrix uniformly distributed in the Circular Orthogonal Ensemble (COE), which is the quotient space $\mathcal{U}(M)/\mathcal{O}(M)$, where $\mathcal{O}(M)$ is the orthogonal group, with normalized Haar measure. Again, this is true for any fixed value of the energy $E$.

In this case the correlation function $\left\langle \prod_{k=1}^n S_{i_ki_{-k}}S^\dag_{j_kj_{-k}}\right\rangle_{\rm COE(M)}$ will be different from zero if and only if the $j$ labels are a permutation of the $i$ labels. Namely, 
\be\label{e0O} \left\langle \prod_{k=1}^n S_{i_ki_{-k}}S^\dag_{j_kj_{-k}}\right\rangle_{\rm COE(M)} = \sum_{\sigma \in S_{2n}} \Wg^{(2)}_M(\sigma)\delta_\sigma[\iv,\jv],\ee where ${\rm Wg}^{(2)}_M$ is the Weingarten function of the COE.
For example, \be\label{S4} \langle S_{12}S^*_{12}S_{34}S^*_{34}\rangle_{\rm COE(M)}
=\frac{M+2}{M(M+1)(M+3)},\ee and \be\label{S4b}  \langle
S_{12}S^*_{14}S_{34}S^*_{23}\rangle_{\rm COE(M)}=\frac{-1}{M(M+1)(M+3)}.\ee

The general case has an expansion in terms of zonal spherical functions given by \cite{coe,matsucompact}
\be\label{oldweing2} \Wg^{(2)}_M(\sigma)=\frac{2^nn!}{(2n)!}\sum_{\lambda\vdash n}\frac{d_{2\lambda}}{[M+1]^\lambda_{(2)}}\omega_\lambda(\sigma).\ee Here $\omega$ are the irreducible characters of that group, and $[M]^\lambda_{(1)}$ is a generalization of the raising factorial. These concepts are reviewed in the appendix.

Weingarten functions were first studied in physics, \cite{samuel,weing} where they found applications to quantum chaos. \cite{BB,esposti,mello} They later found their way into mathematics \cite{collins,matsucompact,ColSni,ColMat,zuber,Banica} and back into other areas of physics
\cite{scott,pineda,cramer,znidaric}.

\section{Semiclassical approximation}

The semiclassical approach to quantum chaotic scattering has been reviewed in detail before.\cite{muller} We present only a brief sketch.

\subsection{The approximation}

In the semiclassical limit $\hbar\to 0$, $M\to \infty$, the element $S_{ab}$ of the $S$
matrix may be approximated by a sum over trajectories $\gamma$ that start at incoming channel $b$
and end at outgoing channel $a$, \cite{jalabert} $S_{ab}=\frac{1}{\sqrt{T_H}}\sum_{\gamma:b\to a}A_\gamma e^{i\mathcal{S}_\gamma/\hbar}$, where $\mathcal{S}_\gamma(E)$ is the trajectory's action and $A_\gamma$ is related to its stability. The quantity $T_H=M\tau_D$ is called the Heisenberg time.

Within this approach, correlation functions involve multiple sums over trajectories which, for chaotic systems, are strongly fluctuating functions of the energy. Averaging over $E$ and using the standard stationary phase
approximation, one arrives at very restrictive conditions that the trajectories must satisfy in order to provide constructive interference. These restrictions are satisfied in the presence of crossings, analogously to what happens in closed systems.\cite{scripta,R2c} 

The theory is perturbatively formulated  in terms of diagrams, as a power series in $1/M$, with the contribution of a diagram being proportional to $M^{V-E}$, where $V$ and $E$ are the numbers of vertices and edges of the diagram, respectively, and with a coefficient which is a rational function of $\epsilon$ (although most advances have happened for in the far simpler case of $\epsilon=0$\cite{BK3,BK4,EPL}).

\subsection{The matrix integrals}

As shown in our previous work,\cite{previous,novaes2,novaes1} the perturbative diagrammatics of the semiclassical approximation can be modelled by appropriate matrix integrals. For ease of notation, let $Z_{\vec i,\vec{j}}=\prod_{k=1}^n Z_{i_k,j_k}$.

For systems with broken time-reversal symmetry, the correlation functions that interest us, Eq. (\ref{corr1}), are given by
\be C_1(\vec{i},\vec{j},\vec{p},\vec{q},\epsilon,M)= \lim_{N\to 0}G_1(\vec{i},\vec{j},\vec{p},\vec{q},\epsilon,M,N), \ee  where \be G_1(\vec{i},\vec{j},\vec{p},\vec{q},\epsilon,M,N)=\frac{1}{\mathcal{Z}_1}\int dZ e^{-M\sum_{q\ge 1}\frac{(1-iq\epsilon)}{q}\tr[(ZZ^\dag)^q]}Z_{\vec i,\vec j}Z^\dag_{\vec p,\vec q}.\ee Here $Z$ and $Z^\dag$ are $N$-dimensional complex matrices and the quantity $\mathcal{Z}_1=\int dZ e^{-M(1-i\epsilon)\tr(ZZ^\dag)}$ is a normalization constant. Notice that the limit $N\to 0$ must be taken after the integral has been performed and expressed as an analytic function of $N$.

For systems with intact time-reversal symmetry, the theory is slightly more complicated. First, we must compute a matrix integral, 
\be G_2(\iv,\epsilon,M,N,W)=\frac{1}{\mathcal{Z}_2}\int dZ e^{-\frac{M}{2}\sum_{q\ge 1}\frac{(1-iq\epsilon)}{q}\tr[(ZZ^T)^q]} \prod_{k=1}^{n}R_{i_k,i_k}R_{i_{-k},i_{-k}}, \ee where now $Z$ and $Z^T$ are $N$-dimensional real matrices and the normalization constant is $\mathcal{Z}_2=\int dZ e^{-\frac{M}{2}(1-i\epsilon)\tr(ZZ^T)}$. Matrix $R$ is given by $R=WQZQ^TW^\dag$, where $W$ is a complex $M\times M$ matrix and $Q$ is a $M\times N$ matrix 
with elements $Q_{ij}=\delta_{ij}$. 

Once the above integral has been computed, it will be a polynomial in the matrix elements of $W$. We must extract from it the coefficient of a particular combination of such elements. In terms of $K=WW^T$, we need the coefficient of $K_{\iv}K^\dag_{\jv}$, where $K_{\iv}=\prod_{k=1}^nK_{i_k,i_{-k}}$. Using the
notation $[x]f$ for the coefficient of $x$ in $f$, we have 
\be C_2(\iv, \jv,\epsilon,M)= \lim_{N\to 0}\left[K_{\iv}K^\dag_{\jv}\right]G_2(\iv,\epsilon,M,N,W).\ee 

\subsection{The normalization constants}

The normalization constants $\Z_1$ and $\Z_2$ are given by \be \Z_\alpha=\int dZ e^{\frac{M}{\alpha} (1-i\epsilon)\tr(ZZ^\dag)},\ee where $Z^\dag=Z^T$ if $Z$ is real. They can be computed by introducing the singular value decomposition $Z=UDV$, where $D$ is real and positive diagonal matrix, while the matrices of left and right singular vectors, $U$ and $V$, are elements of the orthogonal group if $Z$ is real and of the unitary group if $Z$ is complex. Let us unify the treatment by denoting $\U_1(N):=\U(N)$ and $\U_2(N):=\mathcal{O}(N)$. Let $d_\alpha U$ be the Haar measure on $\U_\alpha(N)$. 

The singular value decomposition is in general not unique, since the triple $(UW,D,W^\dag V)$ is equivalent to $(U,D,V)$ for any diagonal $W\in\U_\alpha(N)$. We can make it unique by choosing $UW$ to have real positive elements along the diagonal. The random distribution of $Z$ then implies that $V$ is uniformly distributed in $\U_\alpha(N)$ (with respect to Haar measure), while $U$ is uniformly distributed in the coset space $\U_\alpha(N)/[\U_\alpha(1)]^N$. We normalize the total volume associated with the variables $(U,V)$, \be 
\int_{\U_\alpha(N)}d_\alpha V\int_{\U_\alpha(N)/[\U_\alpha(1)]^N}d_\alpha U=1.\ee

The Jacobian of the singular value decomposition is known in terms of the variables $x_i$, $1\leq i\leq N$, the eigenvalues of $X=D^2$. It is given by \cite{morris,shen,edelman} \be
dZ=d_\alpha Ud_\alpha V\det(X)^{(1-\alpha)/\alpha}|\Delta(X)|^{2/\alpha}dx,\ee where $dx=dx_1\cdots dx_N$ and the factor $\Delta(X)$ is the usual Vandermonde antisymmetric polynomial, \be \Delta(X)=\prod_{i<j}(x_j-x_i).\ee

We therefore have \be \Z_\alpha=\int_0^\infty e^{\frac{M}{\alpha} (1-i\epsilon)\tr(X)}\det(X)^{(1-\alpha)/\alpha}|\Delta(X)|^{2/\alpha}dx.\ee This integral can be computed explicitly because it is a limit case of the Selberg integral. The result is 
\be \Z_\alpha=\left(\frac{\alpha}{M(1-i\epsilon)}\right)^{N^2/\alpha}\prod_{j=1}^N \frac{\Gamma(1+j/\alpha)\Gamma(j/\alpha)}{\Gamma(1+1/\alpha)}.\ee

Notice that taking $b=M(1-i\epsilon)/\alpha$ in Eq. (\ref{SelbergJack}) we can write 
\be\label{SelbergJackNorm} \frac{1}{\mathcal{Z}_\alpha}\int_0^\infty dx
|\Delta(x)|^{2/\alpha}\det(X)^{(1-\alpha)/\alpha}e^{-\frac{M}{\alpha}(1-i\epsilon)\tr
X}J_\lambda^{(\alpha)}(x)=\left([N]^\lambda_{(\alpha)}\right)^2
\frac{\alpha^{n-|\lambda|}}{[M(1-i\epsilon)]^n}.\ee 

\section{Broken time-reversal symmetry}

\subsection{Angular integrals}

The matrix integral required in this case is \be G_1(\vec{i},\vec{j},\vec{p},\vec{q},\epsilon,M,N)=\frac{1}{\mathcal{Z}_1}\int dZ e^{-M\sum_{q\ge 1}\frac{(1-iq\epsilon)}{q}\tr[(ZZ^\dag)^q]}Z_{\vec i,\vec j}Z^\dag_{\vec p,\vec q}.\ee Just like we did for the normalization constants, we introduce the singular value decomposition $Z=UDV$ and integrate over $U$ and $V$. The angular integrals needed are
\be \int_{\U(N)} dU U_{\vec{i},\vec{a}} U^\dag_{\vec{b},\vec{q}}=\sum_{\xi,\tau \in S_n}\Wg^{(1)}_N(\xi^{-1}\tau)\delta_\tau[\vec{q},\vec{i}]\delta_\xi[\vec{b},\vec{a}]\ee
and
\be \int_{\U(N)} dV V_{\vec{a},\vec{j}} V^\dag_{\vec{p},\vec{b}}=\sum_{\rho,\sigma \in S_n}\Wg^{(1)}_N(\rho^{-1}\sigma)\delta_\sigma[\vec{p},\vec{j}]\delta_\rho[\vec{b},\vec{a}].\ee The sum over $\vec{a}$ and $\vec{b}$ gives, according to Eq.(\ref{power1}),
\be \sum_{\vec{a},\vec{b}=1}^N \delta_\xi[\vec{b},\vec{a}]\delta_\rho[\vec{b},\vec{a}]\prod_{k=1}^n
y_{a_k}y_{b_k}=p_{\xi^{-1}\rho}(x),\ee where $y_i$ are the eigenvalues of $D$ and $x_i=y_i^2$ are the eigenvalues of $X=D^2$.

Moreover, we can use the character expansion of ${\rm Wg}^{(1)}_N$ and the orthogonality of characters to get \be\label{G} G_1(\vec{i},\vec{j},\vec{p},\vec{q},\epsilon,M,N)=\frac{1}{n!}\sum_{\tau,\sigma \in S_n} \sum_{\mu\vdash
n}\frac{d_\mu\chi_\mu(\sigma^{-1}\tau)}{\left([N]^\mu_{(1)}\right)^2}\mathcal{R}_\mu(\epsilon,M,N)\delta_\tau[\vec{q},\vec{i}]\delta_\sigma[\vec{p},\vec{j}],\ee where
$\mathcal{R}_\mu(\epsilon,M,N)$ is the integral over the eigenvalues of $X$,
\be \mathcal{R}_\mu(\epsilon,M,N)=\frac{1}{\Z_1}\int_0^\infty dx
e^{-M\sum_{q\ge 1}\frac{(1-iq\epsilon)}{q}\tr
X^q}|\Delta(x)|^2J_{\mu}^{(1)}(x),\ee with  $J_{\mu}^{(1)}(x)$ being a Jack polynomial. 

\subsection{Eigenvalue integral}

In our previous work,\cite{previous} we proceeded from this point by summing the infinite series in the exponent, and then writing the exponential itself as an infinite sum over Schur functions. After taking the limit $N\to 0$, we obtained correlation functions as power series in $\epsilon$, whose coefficients are rational functions of $M$.

Presently, we shall follow a different and somewhat complementary approach. We wish to express correlation functions as power series in $1/M$, whose coefficients are rational functions of $\epsilon$. 

To this end, we do not sum the series in the exponent. Instead, we consider the $q=1$ term of the sum separately, and expand the rest of the exponential according to Eq.(\ref{exposum}):
\be\label{exponen} e^{-M\sum_{q\ge 2}\frac{(1-iq\epsilon)}{q}\tr
X^q}=1+\sum_{m=2}^\infty\sum_{\substack{\beta\vdash m\\v_1(\beta)=0}} \frac{(-M)^{\ell(\beta)}}{z_\beta}g_\beta(\epsilon) p_\beta(x),\ee where \be g_\beta(\epsilon)=\prod_{q\in\beta}(1-iq\epsilon)\ee and the condition $v_1(\beta)=0$ means that $\beta$ has no parts equal to 1. The integrand now contains the product $J_{\mu}^{(1)}(x)p_\beta(x)$, which may be written as a sum over Jack polynomials,
\be J_{\mu}^{(1)}(x)p_\beta(x)=n!\sum_{\rho\vdash n}\frac{1}{z_\rho} \theta^{(1)}_\mu(\rho)p_{\beta+\rho}(x)=\frac{m!}{(n+m)!^2}\sum_{\lambda\vdash n+m}d_\lambda^2 \theta^{(1)}_{\lambda\backslash\mu}(\beta)J_\lambda^{(1)}(x),\ee where we have used Eqs. (\ref{J2p}), (\ref{p2J}) and (\ref{skew}). 

Resorting to Eq.(\ref{SelbergJackNorm}), we arrive at 
\be  \frac{\mathcal{R}_\mu}{\left([N]^\mu_{(1)}\right)^2}=\frac{1}{[M(1-i\epsilon)]^{n}}\left(1+\sum_{m=2}^\infty\sum_{\substack{\beta\vdash m \\ v_1(\beta)=0}} \frac{m!(-M)^{\ell(\beta)}g_\beta(\epsilon)}{z_\beta(n+m)!^2[M(1-i\epsilon)]^{m}}Q_{\mu,\beta}(N)\right),\ee where \be Q_{\mu,\beta}(N)=\sum_{\lambda\vdash n+m}\left(\frac{[N]^\lambda_{(1)}}{[N]^\mu_{(1)}}\right)^2d_\lambda^2 \theta^{(1)}_{\lambda\backslash\mu}(\beta).\ee

\subsection{Final result}

We must now let $N\to 0$. We have seen that $[N]^\lambda_{(1)}\sim t_1(\lambda)N^{D_1(\lambda)}$, where $D_1(\lambda)$ is the size of the Durfee square of $\lambda$. The skew character $\theta^{(1)}_{\lambda\backslash\mu}$ is different from zero only if $\lambda\supset \mu$, which implies $D_1(\lambda)\ge D_1(\mu)$. Therefore, we have 
\be\label{limit} \lim_{N\to 0}\left(\frac{[N]^\lambda_{(1)}}{[N]^\mu_{(1)}}\right)^2\theta^{(1)}_{\lambda\backslash\mu}(\beta)=\left\{\begin{array}{r@{\quad}cr} \left(\frac{t_1(\lambda)}{t_1(\mu)}\right)^2\theta^{(1)}_{\lambda\backslash\mu}(\beta),& \;\mathrm{if}\; D_1(\lambda)= D_1(\mu);\\0,& \;\mathrm{if}\;D_1(\lambda)>D_1(\mu).\end{array}\right.\ee 

Collecting these results, we arrive at 
\be\label{C1final} C_1(\vec{i},\vec{j},\vec{p},\vec{q},\epsilon,M)=\sum_{\tau,\sigma \in S_n} \mathcal{W}^{(1)}_{M,\epsilon}(\sigma^{-1}\tau)\delta_\tau[\vec{q},\vec{i}]\delta_\sigma[\vec{p},\vec{j}],\ee where 
\be\label{newweing} \mathcal{W}^{(1)}_{M,\epsilon}(\pi)=\frac{1}{[M(1-i\epsilon)]^{n}n!}\sum_{\mu\vdash
n}d_\mu\chi_\mu(\pi)\left(1+\sum_{m=2}^\infty\sum_{\substack{\beta\vdash m \\ v_1(\beta)=0}}\frac{E_{\mu,\beta}(\epsilon)}{M^{r(\beta)}}\right),\ee with $r(\beta)=m-\ell(\beta)$ and 
\be\label{E1} E_{\mu,\beta}(\epsilon)=\frac{g_\beta(\epsilon)}{(1-i\epsilon)^m}\frac{m!(-1)^{\ell(\beta)}}{z_\beta(n+m)!^2t_1(\mu)^2}\sum_{\lambda\vdash n+m}d_\lambda^2t_1(\lambda)^2\theta^{(1)}_{\lambda\backslash\mu}(\beta)\delta_{D_1(\lambda),D_1(\mu)}.\ee

We see that the correlation for finite $\epsilon$ has a similar structure to the $\epsilon=0$ one, Eq. (\ref{e0U}), except now the usual Weingarten function is replaced by the generalization $\mathcal{W}^{(1)}_{M,\epsilon}$.  

For example, when the argument is the transposition $(12)$, we now have
\be \mathcal{W}^{(1)}_{M,\epsilon}((12))=\frac{2i\epsilon-1}{(1-i\epsilon)^4M^3}
+\frac{(4i\epsilon-1)(-8\epsilon^2-2i\epsilon+1)}{(1-i\epsilon)^8M^5}+O\left(\frac{1}{M^{7}}\right),\ee which is to be compared to the corresponding series for Eq.(\ref{ex1}), which is 
\be \Wg^{(1)}_M((12))=-\frac{1}{M^3}-\frac{1}{M^5}+O\left(\frac{1}{M^{7}}\right).\ee

\subsection{Recovering $\epsilon=0$}

Naturally, we must have $\mathcal{W}^{(1)}_{M,0}(\pi)={\rm Wg}_M^{(1)}(\pi)$. In the regime $M\gg 1$ this follows simply from the fact that $[M]^\mu_{(1)}\sim M^n$. 

For finite $M$, it is not so easy to see how this equivalence comes about. Comparing Eq. (\ref{newweing}) with Eq. (\ref{oldweing}), we arrive at the non-trivial identity
\be 1+\sum_{m\ge 2}\sum_{\substack{\beta\vdash m \\ v_1(\beta)=0}}\frac{E_{\mu,\beta}(0)}{M^{r(\beta)}}=\frac{M^n}{[M]^\mu_{(1)}},\ee that must be valid for any $\mu$. Unfortunately, although we have checked it numerically to the extent possible, we were not able to prove this.

The main difficulty in understanding these identities, even in the simplest case when $\pi$ is a single cycle (and hence both $\mu$ and $\lambda$ must be hooks) is the rather poor current knowledge about the skew characters $\theta^{(1)}_{\lambda\backslash\mu}(\beta)$. 

Let us mention that, by writing \be \frac{1}{[M]^\mu_{(1)}}=\frac{1}{M^n}\prod_{\square\in \mu}\frac{1}{1+c(\square)/M}=\frac{1}{M^n}\sum_{s=0}^\infty \frac{(-1)^s}{M^s}h_s(A_\mu),\ee where $h_s$ are the complete symmetric polynomials and $A_\mu=\{c(\square),\square \in\mu\}$ is the so-called `content alphabet' of $\mu$, it is possible to obtain separate identities satisfied by the coefficients of different powers of $M$:
\be \sum_{m\ge 2}\sum_{\substack{\beta\vdash m \\ v_1(\beta)=0\\r(\beta)=s}}E_{\mu,\beta}(0)=(-1)^sh_s(A_\mu).\ee Notice that the left-hand-side above is \emph{not} an infinite sum, because of the simultaneous restrictions $v_1(\beta)=0$ and $r(\beta)=s$.

\section{Preserved time-reversal symmetry}

\subsection{Angular integrals}

The matrix integral involved in this case is \be G_2(\iv,\epsilon,M,N,W)=\frac{1}{\mathcal{Z}_2}\int dZ e^{-\frac{M}{2}\sum_{q\ge 1}\frac{(1-iq\epsilon)}{q}\tr[(ZZ^T)^q]}\prod_{k=1}^{n}R_{i_k,i_k}R_{i_k,i_{-k}}.\ee Just like we did in the previous Section, we introduce the singular value decomposition $Z=UDV$, but matrices $U$ and $V$ and now integrated over the orthogonal group. Remembering that $R_{ii}=\sum_{a,b,c,d=1}^MW_{ia}U_{ab}D_bV_{bc}(W^\dag)_{ci}$, the angular integrals needed are
\be \int_{\mathcal{O}(N)} dU U_{\av,\bv}=\sum_{\xi,\tau \in \mathcal{M}_{n}}\Wg^{(O)}_N(\xi^{-1}\tau)\Delta_\tau[\av]\Delta_\xi[\bv]\ee
and
\be \int_{\mathcal{O}(N)} dV V_{\bv,\cv}=\sum_{\rho,\sigma \in \mathcal{M}_{n}}\Wg^{(O)}_N(\rho^{-1}\sigma)\Delta_\rho[\bv]\Delta_\sigma[\cv],\ee where ${\rm Wg}^{(O)}_N=\Wg^{(2)}_{N-1}$ is the Weingarten function of the orthogonal group\cite{o1,o2} and $\Delta_\xi$ was defined in Eq. (\ref{Delta}).

We shall use
\be \sum_{\bv=1}^N \Delta_\xi[\bv]\Delta_\rho[\bv]\prod_{k=1}^n
y_{b_k}y_{b_{-k}}=p_{\xi^{-1}\rho}(x),\ee where $y_{b_k}=y_{b_{-k}}$ are the eigenvalues of $D$ and $x_i=y_i^2$ are the eigenvalues of $X=D^2$. We also need to calculate
\be \left[K_{\iv}\right] \sum_{\av=1}^N \Delta_\tau[\av]\prod_{k=1}^nW_{i_k,a_{k}}W_{i_{-k},a_{-k}}.\ee This is zero unless $\tau$ is the identity, in which case it is one. Also, since $K$ is symmetric,
\be \left[K^{\dag}_{\jv}\right] \sum_{\cv=1}^N \Delta_\sigma[\cv]\prod_{k=1}^nW^{\dag}_{c_k,i_{k}}W^{\dag}_{c_{-k},i_{-k}}\ee is different from zero if and only if $\jv$ and $\sigma(\iv)$ differ by a hyperoctahedral permutation, i.e. it gives rise to a factor \be \sum_{\gamma \in H_n} \delta_\gamma[\jv,\sigma(\iv)].\ee 

Using the zonal spherical function expansion of ${\rm Wg}^{(O)}_N$ and the orthogonality of these functions, we get \be\label{G2} \left[K_{\iv}K^{\dag}_{\jv}\right]G_2(\iv,\epsilon,M,N,W)=\frac{2^nn!}{(2n)!}\sum_{\sigma \in S_{2n}} \sum_{\mu\vdash
n}\frac{d_{2\mu}\omega_\mu(\sigma)}{\left([N]^\mu_{(2)}\right)^2}\mathcal{R}_\mu(\epsilon,M,N)\delta_\sigma[\iv,\jv],\ee where we have combined the sum over $\sigma\in\mathcal{M}_n$ and the sum over $\gamma\in H_n$ into a single sum over $\sigma\in S_{2n}$.
The quantity $\mathcal{R}_\mu(\epsilon,M,N)$ is the integral over the eigenvalues of $X$,
\be \mathcal{R}_\mu(\epsilon,M,N)=\frac{1}{\Z_2}\int_0^\infty dx
e^{-\frac{M}{2}\sum_{q\ge 1}\frac{(1-iq\epsilon)}{q}\tr
X^q}|\Delta(x)|J_{\mu}^{(2)}(x).\ee

\subsection{Eigenvalue integration and final result}

We proceed in the same fashion as in the case of broken time-reversal symmetry. Namely, we use the expansion in Eq.(\ref{exponen}). The integral to be performed now involves the function
\be J_{\mu}^{(2)}(x)p_\beta(x)=\frac{2^{m}m!}{(2n+2m)!}\sum_{\lambda\vdash n+m}d_{2\lambda} \theta^{(2)}_{\lambda\backslash\mu}(\beta)J_\lambda^{(2)}(x),\ee where we have used Eqs. (\ref{J2p2}), (\ref{p2J2}) and (\ref{skew}). Resorting to Eq.(\ref{SelbergJackNorm})
we arrive at 
\be  \frac{\mathcal{R}_\mu}{\left([N]^\mu_{(2)}\right)^2}=\frac{1}{[M(1-i\epsilon)]^{n}}\left[1+\sum_{m=2}^\infty\sum_{\substack{\beta\vdash m \\v_1(\beta)=0}} \frac{2^{r(\beta)}g_\beta(\epsilon)}{M^{r(\beta)}(1-i\epsilon)^m}\frac{(-1)^{\ell(\beta)}m!}{z_\beta(2n+2m)!}\widetilde{Q}_{\mu,\beta}(N)\right],\ee where \be \widetilde{Q}_{\mu,\beta}(N)=\sum_{\lambda\vdash n+m}d_{2\lambda} \theta^{(2)}_{\lambda\backslash\mu}(\beta)\left(\frac{[N]^\lambda_{(2)}}{[N]^\mu_{(2)}}\right)^2.\ee

The limit $N\to 0$ now gives  
\be\label{limit2} \lim_{N\to 0}\left(\frac{[N]^\lambda_{(2)}}{[N]^\mu_{(2)}}\right)^2\theta^{(2)}_{\lambda\backslash\mu}(\beta)=\left\{\begin{array}{r@{\quad}cr} \left(\frac{t_2(\lambda)}{t_2(\mu)}\right)^2\theta^{(2)}_{\lambda\backslash\mu}(\beta),& \;\mathrm{if}\; D_2(\lambda)= D_2(\mu);\\0,& \;\mathrm{if}\;D_2(\lambda)>D_2(\mu);\end{array}\right.,\ee leading to 
\be\label{C2final} C_2(\iv,\jv,\epsilon,M)=\sum_{\sigma \in S_n} \mathcal{W}^{(2)}_{M,\epsilon}(\sigma)\delta_\sigma[\iv,\jv],\ee where 
\be\label{newweing2} \mathcal{W}^{(2)}_{M,\epsilon}(\pi)=\frac{2^nn!}{(2n)![M(1-i\epsilon)]^{n}}\sum_{\mu\vdash
n}d_{2\mu}\omega_\mu(\pi)\left(1+\sum_{m=2}^\infty\sum_{\substack{\beta\vdash m \\ v_1(\beta)=0}}\frac{2^{r(\beta)}\widetilde{E}_{\mu,\beta}(\epsilon)}{M^{r(\beta)}}\right),\ee with $r(\beta)=m-\ell(\beta)$ and 
\be\label{E2} \widetilde{E}_{\mu,\beta}(\epsilon)=\frac{g_\beta(\epsilon)}{(1-i\epsilon)^m}\frac{(-1)^{\ell(\beta)}m!}{z_\beta(2n+2m)!t_2(\mu)^2}\sum_{\lambda\vdash n+m}d_{2\lambda}t_2(\lambda)^2\theta^{(2)}_{\lambda\backslash\mu}(\beta)\delta_{D_2(\lambda),D_2(\mu)}.\ee

We see that, also for intact TRS, the correlation function for finite $\epsilon$ have a similar structure to the $\epsilon=0$ one, Eq. (\ref{e0O}), except for the fact that the usual Weingarten function is replaced by the generalization $\mathcal{W}^{(2)}_{M,\epsilon}$. 

For example, we now have
\be \mathcal{W}^{(2)}_{M,\epsilon}((1)(2))=\frac{1}{(1-i\epsilon)^2M^2}
+\frac{(4i\epsilon-2)}{(1-i\epsilon)^4M^3}+O\left(\frac{1}{M^{4}}\right),\ee which is to be compared to the corresponding series for Eq.(\ref{S4}), which is 
\be \Wg^{(2)}_M((1)(2))=\frac{1}{M^2}-\frac{2}{M^3}+O\left(\frac{1}{M^{4}}\right).\ee

Recovering the usual Weingarten function from $\mathcal{W}^{(2)}_{M,\epsilon}$ when $\epsilon=0$ is again not a trivial procedure, mainly due to the presence of the poorly understood skew Jack character. It boils down to the fact that the identity
\be 1+\sum_{m=2}^\infty\sum_{\substack{\beta\vdash m \\ v_1(\beta)=0}}\frac{2^{r(\beta)}\widetilde{E}_{\mu,\beta}(0)}{M^{r(\beta)}}=\frac{M^n}{[M+1]^\mu_{(2)}}\ee must hold for every $\mu$. Again, we have checked this in some instances but were not able to prove it.

\section{Trace correlations}

In this section we consider a by-product of the matrix elements correlations, which are the trace correlations. Namely, take $\lambda$ a partition of some positive integer, $\lambda\vdash n$, and define 
\be\label{trace} C^{(\alpha)}_{\lambda}(\epsilon,M)=\frac{1}{M^{\ell(\lambda)}}\left\langle \prod_{k=1}^{\ell(\lambda)} \tr\left[(S(E_+)S^\dag(E_-))^{\lambda_i}\right]\right\rangle,\ee where $\alpha=1$ for broken TRS and $\alpha=2$ for preserved TRS.

\subsection{Broken TRS}

To produce $C_{\lambda}(\epsilon,M)$ from our broken-TRS general correlations, Eq. (\ref{C1final}), we take $\vec{p}=\vec{j}$ and $\vec{q}=\pi(\vec{i})$ for some permutation $\pi\in S_n$ having cycle type $\lambda$, and then sum over $\vec{i}$ and $\vec{j}$.

Using \be \sum_{\vec{i},\vec{j}=1}^M\delta_\tau[\vec{\pi(i)},\vec{i}]\delta_\sigma[\vec{j},\vec{j}]=M^{\ell(\sigma)+\ell(\tau\pi)}=p_\sigma(1^M)p_{\tau\pi}(1^M)\ee and Eq. (\ref{p2J}), we arrive at 
\be C^{(1)}_{\lambda}(\epsilon,M)=\frac{1}{M^{n+\ell}(1-i\epsilon)^{n}n!}\sum_{\mu\vdash
n}d_\mu\chi_\mu(\lambda)\left([M]^\mu_{(1)}\right)^2\left(1+\sum_{m=2}^\infty\sum_{\substack{\beta\vdash m \\ v_1(\beta)=0}}\frac{E_{\mu,\beta}(\epsilon)}{M^{r(\beta)}}\right),\ee with $E_{\mu,\beta}(\epsilon)$ given by Eq. (\ref{E1}).
Using a computer, we get: 
\be C^{(1)}_{(1,1)}=\frac{1}{(1-i\epsilon)^2}-\frac{\epsilon^2(4-\epsilon^2)}{M^2(1-i\epsilon)^6}-\frac{\epsilon^2(4+64i\epsilon-85\epsilon^2-28i\epsilon^3+8\epsilon^4)}{M^4(1-i\epsilon)^{10}}+O\left(\frac{1}{M^6}\right),\ee
\be C^{(1)}_{(2,1)}=-\frac{1-2i\epsilon-2\epsilon^2}{(1-i\epsilon)^5}-\frac{\epsilon^2(9+6i\epsilon-21\epsilon^2-2i\epsilon^3+4\epsilon^4)}{M^2(1-i\epsilon)^{9}}+O\left(\frac{1}{M^4}\right),\ee
\be C^{(1)}_{(1,1,1)}=-\frac{1}{(1-i\epsilon)^3}-\frac{3\epsilon^2(3-\epsilon^2)}{M^2(1-i\epsilon)^7}+O\left(\frac{1}{M^6}\right).\ee 

To the best of my knowledge, this is the first time correlation functions like these have been found explicitly. Previously, only the case $\ell(\lambda)=1$ had been considered. In that case, we can go a little further than current results (the most recent ones can be found from the generating functions provided by Borkolaiko and Kuipers\cite{BK2}). For instance, at order $1/M^6$ the function $C_{(1)}(\epsilon,M)$ is given by 
\be C^{(1)}_{(1)}(\epsilon,M)=\cdots -\frac{\epsilon^2(1+72i\epsilon-528\epsilon^2-704i\epsilon^3+180\epsilon^4)}{M^6(1-i\epsilon)^{13}}+\cdots\ee At order $1/M^4$, the function $C_{(2)}(\epsilon,M)$ is given by 
\be C^{(1)}_{(2)}(\epsilon,M)=\cdots -\frac{\epsilon^2(4+120i\epsilon-523\epsilon^2-590i\epsilon^3+216\epsilon^4+32i\epsilon^5)}{M^4(1-i\epsilon)^{12}}+\cdots\ee At order $1/M^2$, the function $C_{(3)}(\epsilon,M)$ is given by 
\be C^{(1)}_{(3)}(\epsilon,M)=\cdots -\frac{\epsilon^2(9+36i\epsilon-75\epsilon^2-84i\epsilon^3+49\epsilon^4+16i\epsilon^5-\epsilon^6)}{M^2(1-i\epsilon)^{11}}+\cdots\ee

\subsection{Intact TRS}

To produce $C_{\lambda}(\epsilon,M)$ from our intact-TRS general correlations, Eq. (\ref{C2final}), we now take $\jv=\pi(\iv)$ for some permutation $\pi\in S_{2n}$ having coset type $\lambda$, and then sum over $\iv$.

Using \be \sum_{\iv=1}^M\delta_\sigma[\iv,\pi(\iv)]=M^{\ell(\sigma\pi)}=p_{\sigma\pi}(1^M)\ee and again Eq. (\ref{p2J}) (notice that we must expand $p$ in terms of $J^{(1)}$ and not $J^{(2)}$, because it depends on the cycle type of $\sigma\pi$, not its coset type), we arrive at \be\label{C2} C^{(2)}_{\lambda}(\epsilon,M)=\frac{2^{n}n!}{(2n)!M^{n+\ell}(1-i\epsilon)^{n}}\sum_{\mu\vdash
n}d_{2\mu}\omega_\mu(\lambda)[M]_{(1)}^{2\mu}\left(1+\sum_{m=2}^\infty\sum_{\substack{\beta\vdash m \\v_1(\beta)=0}} \frac{2^{r(\beta)}\widetilde{E}_{\mu,\beta}(\epsilon)}{M^{r(\beta)}}\right),\ee where $\widetilde{E}_{\mu,\beta}(\epsilon)$ is given in Eq. (\ref{E2}).

In the simplest case $\lambda=(n)$, both $\mu$ and $\lambda$ must be $2$-hooks, according to Eq. (\ref{2hooks}), and the expressions simplify slightly. For the first few values of $n$ and up to the first two orders in $1/M$, they can be obtained from the generating functions provided by Berkolaiko and Kuipers.\cite{BK2} As examples, we can mention
\be C_1^{(2)}(\epsilon,M)=\frac{1}{1-i\epsilon}-\frac{\epsilon^2}{(1-i\epsilon)^3M}+O\left(\frac{1}{M^2}\right),\ee
\be C_2^{(2)}(\epsilon,M)=\frac{1-2i\epsilon-2\epsilon^2}{(1-i\epsilon)^4}-\frac{4\epsilon^2(1-\epsilon^2)}{(1-i\epsilon)^6M}+\left(\frac{1}{M^2}\right).\ee

Our Eq. (\ref{C2}) agrees with those results, but does not really allow us to go any further due to the presence of the rather poorly understood skew Jack character $\theta^{(2)}_{\lambda\backslash\mu}(\beta)$. We can however present the following new result: \be C_{(1,1)}^{(2)}(\epsilon,m)=\frac{1}{(1-i\epsilon)^2}-\frac{2\epsilon^2}{M(1-i\epsilon)^4}+O\left(\frac{1}{M^2}\right).\ee

\section{Conclusion}

Working within a semiclassical approximation, and relying on its matrix integral formulation, we have obtained explicit formulas for some correlation functions related to the energy-dependence of the scattering matrix in chaotic systems. These formulas involve infinite sums over characters and zonal spherical functions of permutation groups. They can be seen as a one-parameter generalization of Weingarten functions from the unitary group $\mathcal{U}(M)$ and the symmetric space $\mathcal{U}(M)/\mathcal{O}(M)$.

Even though our results go beyond what was previously known, they are too involved to provide explicit formulas for high correlations. The main obstacle is our current poor understanding of an important concept in the interface of representation theory and combinatorics: the skew Jack characters. In particular, even recovering the usual $\epsilon=0$ case from our general results is not an easy task, as it implies that some identities have to be satisfied by the skew Jack characters. We could verify these identities in particular cases, but we could not prove them in general.

Skew Jack characters have been receiving some attention recently,\cite{jack1,jack2,jack3,jack4} and we may hope that, with further advances in the mathematical side, the path will eventually be unlocked to access energy-dependent $S$-matrix correlations in more detail. 

\section*{Acknowledgments}

This work enjoyed financial support from FAPEMIG (APQ-00393-14) and from CNPq (PQ-303634/2015-4) I am grateful to Richard Stanley for answering some questions of mine on the MathOverflow platform.

\section{Appendix: Review of known facts}

\subsection{Partitions}

A \emph{partition} is a weakly decreasing sequence of positive integers, $\lambda=(\lambda_1,\lambda_2,\ldots)$. The number of non-zero parts is its length, $\ell(\lambda)$. By $\lambda\vdash n$ or $|\lambda|=n$ we mean $\sum_{i=1}^{\ell(\lambda)}\lambda_i=n$. So $(3,2,2,1)\vdash 8$; we also use the notation $(3,2^2,1)\equiv(3,2,2,1)$. The rank of a partition is defined as $r(\lambda)=|\lambda|-\ell(\lambda)$. Partitions of the form \be (k,\underbrace{1,\ldots,1}_{n-k \text{ times}})= (k,1^{n-k})\ee are called \emph{hooks}.

Partitions are usually identified with so-called Young (or Ferrers) \emph{diagrams}. A partition $\lambda$ is seen as a collection of boxes arranged in left-justified rows, with row $i$ having $\lambda_i$ boxes. We say that $\lambda$ `covers' $\mu$, $\lambda\supset \mu$, if $\lambda_i\ge\mu_i$ for all $i$. This is true if the diagram of $\lambda$ literally covers the diagram of $\mu$. The object $\lambda\backslash\mu$, called a \emph{skew diagram}, is then defined to be the collection of those boxes in $\lambda$ that do not belong to $\mu$.

If a box occupies the $j$th position in the $i$th row, we say it has coordinates $(i,j)$. Its \emph{content} is defined as $c(\square)=j-i$. This can be generalized to the $\alpha$-content, which is given by $c_\alpha(\square)=\alpha(j-1)-i+1$. Define the total multiplicative $\alpha$-content as 
\be t_\alpha(\lambda)=\prod_{\square \in \lambda}c_\alpha(\square),\ee where the product includes only those boxes whose $\alpha$-content is not zero. 

In the diagrams below we show the partition $(4^2,2^2,1)$, and we have filled each box with its $1$-content on the left and with its $2$-content on the right:
\begin{figure}[hb]
\includegraphics[scale=0.3,clip]{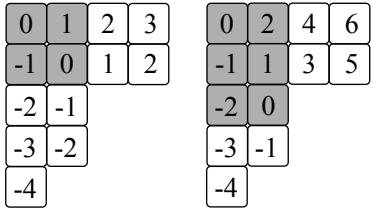}
\caption{Diagram of the partition $(4^2,2^2,1)$, showing 1-contents on the left and 2-contents on the right. We highlighted the Durfee square on the left and the Durfee 2-rectangle on the right.}
\end{figure}

The \emph{Durfee square} of $\lambda$ is the largest square collection of boxes that is covered by $\lambda$. Let $D(\lambda)$ denote the number of boxes on the diagonal of the Durfee square of $\lambda$. Equivalently, it is the number of boxes in $\lambda$ with zero content. A partition $\lambda$ is a hook if and only if $D(\lambda)=1$. The Durfee square is marked in gray in the left diagram above.

The Durfee square may be generalized to the Durfee \emph{$\alpha$-rectangle}, the largest rectangle covered by $\lambda$ whose lower-right corner has zero $\alpha$-content. Let $D_\alpha(\lambda)$ be the horizontal size of the Durfee $\alpha$-rectangle of $\lambda$, i.e. the number of boxes in $\lambda$ with zero $\alpha$-content. The Durfee $2$-rectangle is marked in gray in the right diagram above. We may call a partition $\lambda$ with $D_\alpha(\lambda)=1$ an $\alpha$-hook; for $\alpha=2$ these are of the form $(k_1,k_2,1^{n-k_1-k_2})$.

\subsection{Permutation groups}

Let $S_n$ be the group of all permutations acting on the set $[n]:=\{1,...,n\}$. To a given permutation $\pi\in S_n$ we associate its \emph{cycle type}: a partition of $n$ whose parts are the lengths of the cycles of $\pi$. Permutation $(1\,2\cdots n)$ has cycle type $(n)$, while the identity permutation has cycle type $(1^n)$. The conjugacy class $\mathcal{C}_\lambda$ contains all permutations with cycle type $\lambda$, and its size is $|\mathcal{C}_\lambda|=\frac{n!}{z_\lambda},$ where \be z_\lambda=\prod_j j^{v_j}v_j!,\ee with $v_j(\lambda)$ being the number of times part $j$ appears in $\lambda$.

Irreducible representations (irreps) of $S_n$ are also labelled by partitions of $n$, and we denote by $\chi_\lambda(\mu)$ the character of a permutation of cycle type $\mu$ in the irrep labelled by $\lambda$. The quantity $d_\lambda:=\chi_\lambda(1^n)$ is the dimension of such irrep, and is given by \be\label{dimen} d_\lambda=n!\prod_{i=1}^{\ell(\lambda)}\frac{1}{(\lambda_i-i+\ell)!}\prod_{j=i+1}^{\ell(\lambda)}(\lambda_i-\lambda_j+j-i).\ee Another particular case is $\chi_\lambda(n)$, which is different from zero only if $\lambda$ is a hook, and \be\label{chahooks}\chi_{(k,1^{n-k})}(n)=(-1)^{n-k}.\ee

Characters satisfy two orthogonality relations, 
\be \sum_{\mu\vdash
n}\chi_\mu(\lambda)\chi_\mu(\omega)=z_\lambda\delta_{\lambda,\omega}, \quad
\sum_{\lambda\vdash n}\frac{1}{z_\lambda} \chi_\mu(\lambda) \chi_\omega(\lambda)=
\delta_{\mu,\omega}.\ee The latter is generalized as a sum over permutations as \be
\frac{1}{n!}\sum_{\pi\in S_n}\chi_\mu(\pi) \chi_\lambda(\pi\sigma)=
\frac{\chi_\lambda(\sigma)}{d_\lambda}\delta_{\mu,\lambda}.\ee
 
Let $[-n]:=\{-1,...,-n\}$ and let $[n]\cup[-n]$ be a set with $2n$ elements. A \emph{matching} on this set is a collection of $n$ disjoint subsets with two elements each (`blocks'), such as \be\mathfrak{t}:=\{\{1,-1\},\{2,-2\},...,\{n,-n\}\}.\ee The above matching is said to be `trivial'. We denote the set of all matchings on $[n]\cup[-n]$ by $\mathcal{M}_n$, and we will consider the group $S_{2n}$ acting on $\mathcal{M}_n$ as follows: if block $\{a,b\}$ belongs to matching $\mathfrak{m}$, then block $\{\pi(a),\pi(b)\}$ belongs to $\pi(\mathfrak{m})$. 

The \emph{hyperoctahedral group} $H_n\subset S_{2n}$, with $|H_n|=2^nn!$ elements, is the centralizer of $\mathfrak{t}$ in $S_{2n}$, i.e. $H_n=\{h\in S_{2n}, h(\mathfrak{t})=\mathfrak{t}\}$. Elements in the coset $S_{2n}/H_n$ may therefore be represented by matchings. Also important is the double coset $H_n\backslash S_{2n}/H_n$: permutations $\pi$ and $\sigma$ belong to the same double coset if and only if $\pi=h_1\sigma h_2$ for some $h_1,h_2\in H_n$. 

The notion of cycle type is replaced, in this context, by that of \emph{coset type}. Given a matching $\mathfrak{m}$, let
$\mathcal{G}_\mathfrak{m}$ be a graph with $2n$ vertices having labels in $[n]\cup[-n]$, two vertices
being connected by an edge if they belong to the same block in either $\mathfrak{m}$ or $\mathfrak{t}$. Since each vertex belongs to two edges, all connected components of
$\mathcal{G}_\mathfrak{m}$ are cycles of even length. The coset type of $\mathfrak{m}$ is the partition of $n$ whose parts are half the number of edges in the connected components of $\mathcal{G}_\mathfrak{m}$ (see Figure 1). For example, if $\pi$ fixes all points of $[-n]$ and has cycle type $\lambda$ when restricted to $[n]$, or vice-versa, then it has coset type $\lambda$. We denote by $[\lambda]$ the coset type of $\lambda$.

\begin{figure}[bt]
\includegraphics[scale=1,clip]{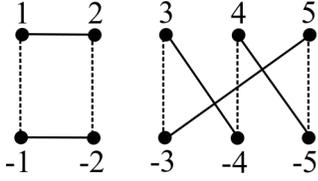}
\caption{The graph associated with matching $\mathfrak{m}=\{\{1,2\},\{-1,-2\},\{3,-4\},\{4,-5\},\{5,-3\}\}$, whose coset type is $(3,2)$.}
\end{figure}

Coset type is clearly invariant under multiplication by hyperoctahedral elements; double cosets are thus labelled by partitions of $n$, so that $\pi$ and $\sigma$ belong to the double coset if and only if they have the same coset type. 

The average \be \omega_\lambda(\tau)=\frac{1}{|H_n|}\sum_{\xi \in
H_n}\chi_{2\lambda}(\tau\xi)\ee is called a zonal spherical function. It is invariant under multiplication by elements of $H_n$ and hence depends only on the coset type of its argument. The simplest case is $\omega_\lambda(1^{n})=1$. These functions satisfy \be\label{omegas} \sum_{\tau\in
\mathcal{M}_{n}}\omega_\lambda(\tau)\omega_\mu(\tau^{-1}\sigma)=\delta_{\lambda,\mu}
\frac{(2n)!}{2^nn!}\frac{\omega_\lambda(\sigma)}{d_{2\lambda}},\ee and also \be
\sum_{\lambda\vdash
n}d_{2\lambda}\omega_\lambda(\sigma)\omega_\lambda(\tau)=\frac{(2n)!}{|H_\tau|}\delta_{[\sigma],[\tau]}.\ee
In particular, $\omega_\lambda(n)$ is different from zero only if $D_2(\lambda)=1$, i.e. if $\lambda$ is a $2$-hook, and in that case it is given by \be\label{2hooks}\omega_\lambda(n)=\frac{t_2(\lambda)}{|H_{n-1}|}.\ee This is an analogue of (\ref{chahooks}).

\subsection{Jack polynomials and Jack characters}

Let $x_i$, $1\leq i\leq N$, be the eigenvalues of matrix $X$. Power sum symmetric functions in these variables can be defined as $p_n(x)=\tr(X^n)$ and generalized as \be p_\lambda(x)=\prod_{i=1}^{\ell(\lambda)}p_{\lambda_i}(x)\ee for $\lambda=(\lambda_1,\lambda_2,...)$ an integer partition. It can be seen from the Taylor series that these functions satisfy the identity \be\label{exposum} e^{\sum_{q=1}^\infty\frac{1}{q}\tr(X^q)}=\sum_{n=0}^\infty\sum_{\lambda\vdash n}\frac{1}{z_\lambda} p_\lambda(x).\ee

Given $\tau\in S_n$ and two sequences, $\vec{j}=(j_1,\ldots,j_n)$ and $\vec{m}=(m_1,\ldots,m_n)$, define the function \be \delta_\tau[\vec{j},\vec{m}]=\prod_{k=1}^n\delta_{j_km_{\tau(k)}},\ee which is equal to 1 if and only if the sequences match up to a the permutation $\tau$. Then we have \be\label{power1} \sum_{\vec{j},\vec{m}=1}^N \delta_\tau[\vec{j},\vec{m}]\delta_\sigma[\vec{j},\vec{m}]\prod_{k=1}^n
y_{j_k}y_{m_k}=p_{\tau^{-1}\sigma}(x),\ee where $x_j=y_j^2$. 

Also, given $\xi\in \mathcal{M}_{n}$ and $\iv=(i_{-n},\ldots,i_{-1},i_1,\ldots,i_n)$, define the function \be\label{Delta} \Delta_\xi[\iv]=\prod_{k=1}^n \delta_{i_{\xi(k)},i_{\xi(-k)}},\ee which is equal to 1 if and only if the elements of the sequence $\iv$ are pairwise equal according to the matching $\xi$. Then we have \be\label{power2} \sum_{\iv=1}^N \Delta_\xi[\iv]\Delta_\rho[\iv]\prod_{k=1}^n y_{i_k}y_{i_{-k}}=p_{\xi^{-1}\rho}(x),\ee where $y_{i_{-k}}=y_{i_k}$ and $x_i=y_i^2$.

Jack polynomials $J_\lambda^{(\alpha)}(x)$ are another family of homogeneous symmetric functions, which depend on a partition and a parameter $\alpha$. The most interesting cases are $\alpha=1$ and $\alpha=2$. In the first case we have
$J_\lambda^{(1)}(x)=\frac{|\lambda|!}{d_\lambda}s_\lambda(x),$ where $s_\lambda$ are the celebrated Schur functions. Out of many interesting properties of these functions, we may mention that they are orthogonal characters of the unitary group $\U(N)$ and therefore satisfy
\be \int_{\U(N)}J_\lambda^{(1)}(U)J_\mu^{(1)}(U^\dag)dU=\frac{d_\lambda^2}{n!^2}\delta_{\lambda,\mu}.\ee By $J_\lambda^{(1)}(U)$ in the above equation we mean the function $J_\lambda^{(1)}$ computed at the eigenvalues of the matrix $U$.

The $\alpha=1$ Jack polynomials are written in terms of power sums as
\be\label{J2p} J_\lambda^{(1)}(x)= \frac{1}{d_\lambda}\sum_{\pi\in
S_n}\chi_\lambda(\pi)p_\pi(x)=\frac{n!}{d_\lambda}\sum_{\mu\vdash
n}\frac{1}{z_\mu}\chi_\lambda(\mu)p_\mu(x).\ee The orthogonality of the permutation group characters allows this expression to be inverted as
\be\label{p2J} p_\mu(x)=\frac{1}{n!}\sum_{\lambda\vdash
n}d_\lambda \chi_\lambda(\mu)J_\lambda^{(1)}(x).\ee 

When $\alpha=2$, the Jack polynomials are called zonal polynomials. Among its interesting properties, they are produced from averages of unitary characters over the orthogonal subgroup,
\be \int_{\mathcal{O}(N)}J_{2\lambda}^{(1)}(AU)dU=\frac{d_\lambda}{n!}\frac{J_\lambda^{(2)}(A^TA)}{J_\lambda^{(2)}(1^N)}.\ee (In the above equation $1^N$ is the identity matrix in $N$ dimensions and $2\lambda=(2\lambda_1,2\lambda_2,...)$.) 
They are related to power sums according to
\be\label{J2p2} J_\lambda^{(2)}(x)= n!\sum_{\mu\vdash
n}\frac{2^{r(\mu)}}{z_\mu} \omega_\lambda(\mu)p_\mu(x),\ee and \be\label{p2J2} p_{\mu}(x)=\frac{2^nn!}{(2n)!}\sum_{\lambda\vdash n}d_{2\lambda}
\omega_\lambda(\mu)J_\lambda^{(2)}(x).\ee 

For general $\alpha$ the so-called \emph{Jack characters}, denoted by $\theta_\lambda^{(\alpha)}(\mu)$, are defined through the relation
\be J_\lambda^{(\alpha)}(x)= n!\sum_{\mu\vdash
n}\frac{\alpha^{r(\mu)}}{z_\mu}\theta^{(\alpha)}_\lambda(\mu)p_\mu(x). \ee Notice that \be \theta_\lambda^{(1)}(\mu)=\frac{\chi_\lambda(\mu)}{d_\lambda},\quad \theta_\lambda^{(2)}(\mu)=\omega_\lambda(\mu).\ee

Define the coefficients $c_{\mu\nu}^{\lambda}(\alpha)$ by the product expansion 
\be J^{(\alpha)}_{\mu}(x)J^{(\alpha)}_\nu(x)=\sum_{\lambda}c_{\mu\nu}^{\lambda}(\alpha)\frac{J^{(\alpha)}_\lambda(x)}{j^{(\alpha)}_\lambda},\ee where \be j^{(\alpha)}_\lambda=2^n n!^2\sum_{\mu\vdash |\lambda|}\frac{2^{r(\mu)}}{z_\mu} \left(\theta^{(\alpha)}_\lambda(\mu)\right)^2=\left\{\begin{array}{r@{\quad}cr} |\lambda|!^2/d_\lambda^2,& \;\mathrm{if}\; \alpha=1;\\|2\lambda|!^2/d_{2\lambda},& \;\mathrm{if}\; \alpha=2.\end{array}\right.\ee These coefficients are used to define Jack polynomials associated with skew diagrams as \be J^{(\alpha)}_{\lambda\backslash\mu}(x)=\sum_{\nu}\frac{c_{\mu\nu}^{\lambda}(\alpha)}{j_\nu^{(\alpha)}}J^{(\alpha)}_\nu(x),\ee and this in turn leads to the definition of \emph{skew Jack characters} through
\be J^{(\alpha)}_{\lambda\backslash\mu}(x)=\sum_{\nu}\frac{|\nu|!\alpha^{r(\nu)}}{z_\nu}\theta^{(\alpha)}_{\lambda\backslash\mu}(\nu)p_\nu(x). \ee These skew characters may be written as a sum over usual characters:
\be\label{skew}\theta^{(\alpha)}_{\lambda\backslash\mu}(\nu)=\frac{|\lambda|!}{|\nu|!}\alpha^{|\lambda|-|\nu|}\sum_{\rho}\frac{|\rho|!\alpha^{r(\rho)}}{z_\rho}\theta^{(\alpha)}_\mu(\rho)\theta^{(\alpha)}_\lambda(\rho+\nu),\ee where $\rho+\nu$ is the partition which contains the union of the parts of $\rho$ and $\nu$.

An important particular value of Jack polynomials is when $x_i=1$ for all $1\le i \le N$. Then \be J^{(\alpha)}_{\lambda}(1^N)=\alpha^{|\lambda|}\prod_{i=1}^{\ell(\lambda)}\frac{\Gamma(\lambda_i+(N-i+1)/\alpha)}{\Gamma((N-i+1)/\alpha)}.\ee Using $\Gamma(z+1)=z\Gamma(z)$ we get 
\be [N]^\lambda_{(\alpha)}:=J^{(\alpha)}_{\lambda}(1^N)=\prod_{i=1}^{\ell(\lambda)}\prod_{j=1}^{\lambda_i}(\alpha(j-1)+N-i+1).\ee In terms of the content notation introduced previously, this is 
\be [N]^\lambda_{(\alpha)}=\prod_{\square\in\lambda}(N+c_\alpha(\square)).\ee

Rather curiously, since the right hand side of the above equation is a polynomial in $N$, it makes sense to let $N\to 0$. It is clear that the smallest power of $N$ in $[N]^\lambda_{(\alpha)}$ is given by the number of boxes in $\lambda$ having zero $\alpha$-content. By definition, this is $D_\alpha(\lambda)$. On the other hand, the coefficient of this power is just the total multiplicative $\alpha$-content. Therefore,
\be [N]^\lambda_{(\alpha)}\sim t_\alpha(\lambda)N^{D_\alpha(\lambda)}\quad (N\to 0).
\ee

Yet another important property of Jack polynomials is that they satisfy nice generalizations of the celebrated Selberg integral.\cite{selberg} The one we shall use is this: 
\be\label{SelbergJack} \int_0^\infty dx
|\Delta(x)|^{2/\alpha}\det(X)^{(1-\alpha)/\alpha}e^{-b\tr
X}J_\lambda^{(\alpha)}(x)=\frac{\left([N]^\lambda_{(\alpha)}\right)^2}{\alpha^{|\lambda|}b^{n+N^2/\alpha}}
\prod_{j=1}^N
\frac{\Gamma(1+j/\alpha)\Gamma(j/\alpha)}{\Gamma(1+1/\alpha)}.\ee


\begin{thebibliography}{99}

\bibitem{micro1} B. Dietz, A. Richter, Chaos {\bf 25}, 097601 (2015).

\bibitem{micro2} S. Kumar, A. Nock, H.-J. Sommers, T. Guhr, B. Dietz, M. Miski-Oglu, A. Richter, and F. Sch\"afer, Phys. Rev. Lett. {\bf 111}, 030403 (2013).

\bibitem{micro3} S. Barkhofen, T. Weich, A. Potzuweit, H.-J. St\"ockmann, U. Kuhl, and M. Zworski, Phys. Rev. Lett. {\bf 110}, 164102 (2013).

\bibitem{micro4} B. Dietz, H.L. Harney, A. Richter, F. Sch\"afer, H.A. Weidenm\"uller,
Physics Letters B {\bf 685}, 263 (2010).

\bibitem{micro5} A. B\"acker, R. Ketzmerick, S. L\"ock, M. Robnik, G. Vidmar, R. H\"ohmann, U. Kuhl, and H.-J. St\"ockmann, Phys. Rev. Lett. {\bf 100}, 174103 (2008).

\bibitem{micro6} S. Hemmady, J. Hart, X, Zheng, T.M. Antonsen Jr., E. Ott and S.M. Anlage, Phys. Rev. B {\bf 74}, 195326 (2006).

\bibitem{elec1} L.A. Ponomarenko, F. Schedin, M.I. Katsnelson, R. Yang, E.W. Hill, K.S. Novoselov, A.K. Geim, Science  {\bf 320}, 356 (2008).

\bibitem{elec2} S. L\"uscher, T. Heinzel, K. Ensslin, W. Wegscheider, and M. Bichler,
Phys. Rev. Lett. {\bf 86}, 2118 (2001).

\bibitem{elec3} S.F. Godijn, S. M\"oller, H. Buhmann, L.W. Molenkamp, and S.A. van Langen, Phys. Rev. Lett. {\bf 82}, 2927 (1999).

\bibitem{elec4} A.M. Chang, H.U. Baranger, L.N. Pfeiffer, and K.W. West,
Phys. Rev. Lett. {\bf 73}, 2111 (1994).

\bibitem{elec5} C.M. Marcus, A.J. Rimberg, R.M. Westervelt, P.F. Hopkins, and A.C. Gossard, Phys. Rev. Lett. {\bf 69}, 506 (1992).

\bibitem{haake} F. Haake, {\it Quantum Signatures of Chaos} (Springer, 2001).

\bibitem{beenakker} C.W.J. Beenakker, Rev. Mod. Phys. {\bf 69}, 731 (1997).

\bibitem{mehta} M.L. Mehta, {\it Random Matrices} (Academic Press, 3rd edition, 2004).

\bibitem{prb78} M. Novaes, Phys. Rev. B {\bf 78}, 035337 (2008).
\bibitem{prb80} B.A. Khoruzhenko, D.V. Savin, and H.J. Sommers, Phys. Rev. B {\bf 80}, 125301 (2009).
\bibitem{prb80mac} F.A.G. Almeida, S. Rodr\'iguez-P\'erez and A.M.S. Mac\^edo, Phys. Rev. B {\bf 80}, 125320 (2009).
\bibitem{prl101} P. Vivo, S.N. Majumdar, O. Bohigas, Phys. Rev. Lett. {\bf 101}, 216809 (2008).

\bibitem{samuel} S. Samuel, J. Math. Phys. {\bf 21}, 2695 (1980).
\bibitem{BB} P.W. Brouwer and C.W.J. Beenakker, J. Math. Phys. {\bf 37}, 4904 (1996).
\bibitem{esposti} M. Degli Esposti and A. Knauf, J. Math. Phys. {\bf 45}, 4957 (2004).
\bibitem{collins} B. Collins, Int. Math. Res. Not. {\bf 17}, 953 (2003).

\bibitem{coe} S. Matsumoto, Random Matrices: Theory Appl. {\bf 1}, 1250005 (2012).
\bibitem{matsucompact} S. Matsumoto, Random Matrices: Theory Appl. {\bf 2}, 1350001 (2013).

\bibitem{sigma1} J.J.M. Verbaarschot, H.A. Weidenm\"uller and M.R. Zirnbauer, Phys. Rep. {\bf 129}, 367 (1985).

\bibitem{sigma2} N. Lehmann, D. V. Savin, V. V. Sokolov and H.-J. Sommers, Physica D {\bf 86}, 572 (1995).

\bibitem{sigma3} Y.V. Fyodorov and H.-J. Sommers, J. Math. Phys. {\bf 38}, 1918 (1997).

\bibitem{new1} J.J.M. Verbaarschot, Ann. Phys. {\bf 168}, 368 (1986).
\bibitem{new2} D.V. Savin, Y.V. Fyodorov, and H.J. Sommers, Acta Phys. Pol. A {\bf 109}, 53 (2006).
\bibitem{new3} E.D. Davis and D. Boos\'e, Phys. Lett. B {\bf 211}, 379 (1988).
\bibitem{new4} E.D. Davis and D. Boos\'e, Z. Phys. A {\bf 332}, 427 (1989).

\bibitem{new5} H.A. Weidenm\"uller, Ann. Phys. {\bf 158}, 120 (1984).

\bibitem{jalabert} R.A. Jalabert, H.U. Baranger and A.D. Stone, Phys. Rev. Lett. {\bf 65}, 2442 (1990).

\bibitem{sieber} K. Richter and M. Sieber, Phys. Rev. Lett. {\bf 89}, 206801 (2002).

\bibitem{muller} S. M\"uller, S. Heusler, P. Braun and F. Haake, New J. Phys. {\bf 9}, 12 (2007).

\bibitem{KS1} J. Kuipers and M. Sieber, Nonlinearity {\bf 20}, 909 (2007).

\bibitem{KS2} J. Kuipers and M. Sieber, Phys. Rev. E {\bf 77}, 046219 (2008).

\bibitem{BK1} G. Berkolaiko and J. Kuipers, J. Phys. A {\bf 43}, 035101 (2010).

\bibitem{andreev1} J. Kuipers, D. Waltner, C. Petitjean, G. Berkolaiko and K. Richter, Phys. Rev. Lett. {\bf 104}, 027001 (2010).

\bibitem{andreev2} J. Kuipers, T. Engl, G. Berkolaiko, C. Petitjean, D. Waltner and K. Richter, Phys. Rev. B {\bf 83}, 195315 (2011).

\bibitem{BK2} G. Berkolaiko and J. Kuipers, New J. Phys {\bf 13}, 063020 (2011).

\bibitem{previous} M. Novaes, J. Math. Phys. {\bf 56}, 062109 (2015).

\bibitem{novaes2} M. Novaes, J. Phys. A {\bf 46}, 502002 (2013).

\bibitem{novaes1} M. Novaes, Ann. Phys. {\bf 361}, 51 (2015).

\bibitem{macdonald} I.G. Macdonald, {\it Symmetric Functions and Hall Polynomials}, 2nd ed. (Oxford University Press, Oxford, 1995).

\bibitem{sagan} B. Sagan, {\it The Symmetric Group: Representations, Combinatorial Algorithms, and Symmetric Functions}, 2nd ed. (Springer, 2001).

\bibitem{weing} D. Weingarten, J. Math. Phys. {\bf 19}, 999 (1978).

\bibitem{mello} P.A. Mello, J. Phys. A: Math. Gen. {\bf 23} 4061 (1990).

\bibitem{ColSni} B. Collins and P. \'Sniady, Commun. Math. Phys. {\bf 264}, 773 (2006).

\bibitem{ColMat} B. Collins and S. Matsumoto, J. Math. Phys. {\bf 50}, 113516 (2009).

\bibitem{zuber} J.-B. Zuber, J. Phys. A: Math. Theor. {\bf 41}, 382001 (2008).

\bibitem{Banica} T. Banica, B. Collins and J.-M. Schlenker, J. Combinat. Theory A {\bf 118}, 78 (2011).

\bibitem{scott} A.J. Scott, J. Phys. A: Math. Theor. {\bf 41}, 055308 (2008).

\bibitem{pineda} M. \v{Z}nidari\v{c}, C. Pineda and I. Garc\'ia-Mata, 
Phys. Rev. Lett. {\bf 107}, 080404 (2011).

\bibitem{cramer} M. Cramer, New J. Phys. {\bf 14}, 053051 (2012).

\bibitem{znidaric} Vinayak and M. \v{Z}nidari\v{c}, J. Phys. A: Math. Theor. {\bf 45},
125204 (2012).

\bibitem{scripta} M. Sieber and K. Richter, Physica Scripta {\bf 2001}, 128 (2001). 

\bibitem{R2c} S. M\"uller, S. Heusler, P. Braun, F. Haake, and A. Altland, Phys. Rev. E {\bf 72}, 046207 (2005).

\bibitem{BK3} G. Berkolaiko and J. Kuipers, J. Math. Phys. {\bf 54}, 112103 (2013).
\bibitem{BK4} G. Berkolaiko and J. Kuipers, J. Math. Phys. {\bf 54}, 123505 (2013).
\bibitem{EPL} M. Novaes, Europhys. Lett. {\bf 98}, 20006 (2012).

\bibitem{morris} T.R. Morris, Nuclear Physics B {\bf 356}, 703 (1991).
\bibitem{shen} J. Shen, Linear Algebra Appl. {\bf 326}, 1 (2001). 
\bibitem{edelman} A. Edelman and N.R. Rao, Acta Numerica {\bf 14},
233 (2005).

\bibitem{o1} B. Collins and P. Sniady, Commun. Math. Phys.
{\bf 264}, 773 (2006).

\bibitem{o2} B. Collins and S. Matsumoto, J. Math. Phys. {\bf 50}, 113516 (2009).

\bibitem{jack1} E. Vassilieva, J. Algebr. Comb. {\bf 42}, 51 (2015).
\bibitem{jack2} M. Do\l ega, V. F\'eray, and P. Sniady, S\'eminaire Lotharingien de Combinatoire {\bf 70}, B70j (2014).
\bibitem{jack3} M. Do\l ega and V. F\'eray, Duke Math. J. {\bf 165}, 1193 (2016).
\bibitem{jack4} P. Sniady, arXiv:1506.06361v2

\bibitem{selberg} P.J. Forrester and S.O. Warnaar, Bull. Amer. Math. Soc. {\bf 45}, 489 (2008).

\end{thebibliography}
\end{document}